\documentstyle[preprint,eqsecnum,prb,aps]{revtex}
\newcommand{\mathrm}{\rm}
\newcommand{\mathcal}{\cal}
\newcommand{\tr}{\mathrm{tr}}
\newcommand{\D}{\mathcal{D}}

\newcommand{\E}{\mathcal{E}}
\renewcommand{\P}{\mathcal{P}}
\renewcommand{\Re}{\mathrm{Re}}

\newcommand{\Q}{\mathcal{Q}}
\begin{document}
\draft
\title{Fully Electrified Neugebauer Spacetimes}
\author{Frederick J.~Ernst}
\address{FJE Enterprises, Rt.~1, Box~246A, Potsdam, NY 13676}
\maketitle
\begin{abstract}
Generalizing a method presented in an earlier paper, we express the
complex potentials $\E$ and $\Phi$ of {\em all} stationary axisymmetric
electrovac spacetimes that correspond to axis data of the form
$$
\E(z,0) = \frac{U-W}{U+W} \; , \; \Phi(z,0) = \frac{V}{U+W} \; ,
$$
where
\begin{eqnarray*}
U & = & z^{2} + U_{1} z + U_{2} \; , \\
V & = & V_{1} z + V_{2} \; , \\
W & = & W_{1} z + W_{2} \; ,
\end{eqnarray*}
in terms of the complex parameters $U_{1}$, $V_{1}$, $W_{1}$, $U_{2}$,
$V_{2}$ and $W_{2}$, that are directly associated with the various
multipole moments.
\end{abstract}

\section{Introduction}

This is the second paper of a short series of papers, all of which are
concerned with those stationary axisymmetric solutions of the Einstein and
the Einstein-Maxwell equations that are characterized by axis data of the
form
\begin{equation}
\E = \frac{U-W}{U+W} \; , \; \Phi = \frac{V}{U+W} \; ,
\end{equation}
with
\begin{mathletters}
\begin{eqnarray}
U & = & \sum_{a=0}^{n} U_{a} z^{n-a} \; , \\
V & = & \sum_{a=1}^{n} V_{a} z^{n-a} \; , \\
W & = & \sum_{a=1}^{n} W_{a} z^{n-a} \; ,
\end{eqnarray}
\end{mathletters}
where $\E$ and $\Phi$ are the complex potentials of Ernst, $z$ is the
Weyl canonical coordinate, and the coefficients in the polynomials are
complex constants.

In the first paper\cite{One} of this series, to which we shall henceforth
refer as Ref.\ 1, we found that it was extraordinarily simple to construct
the general solution of the vacuum problem ($V=0$) for the case $n=2$.
That is, we described a procedure that allows one to express the complex
potential $\E$ and the metrical fields $\omega$ and $\gamma$ of an exact
vacuum solution ($V=0$) directly in terms of the complex parameters $U_{a},
W_{a} \; (a=1,\ldots,n)$ with $U_{0}=1$.  In particular, we showed that
the resulting family of solutions contained as a special case the vacuum
limit of an electrovac solution published recently by Manko et
al.\cite{Manko}

In the present paper, we undertake the extension of the methods of
Ref.~1 to electrovac fields ($V \ne 0$).  This extension is far from
trivial, and to construct it we found that it was necessary to attain
first an understanding of why everything worked out so handily in the
vacuum case.

An additional complexity of the electrovac problem arises from the fact
that, while all the vacuum solutions could be constructed from Minkowski
space by applying successive quadruple-Neugebauer B\"{a}cklund
transformations,\cite{Neu} or double-Harrison B\"{a}cklund
transformations,\cite{Har} there was no single known Kinnersley-Chitre
transformation that could yield all the electrovac solutions with axis
data of the type we are considering.  This forced us to consider what
we call a {\em complexified Cosgrove transformation}, which will be
defined in this paper.

\section{The axis relation}

Using the Hauser-Ernst {\em axis relation},\cite{RetzH,RetzE} one can
always identify a Kinnersley-Chitre transformation that will, in principle,
produce from Minkowski space a spacetime with any specified axis data.
Of course, it may be difficult to solve in closed form the associated
homogeneous Hilbert problem.

\subsection{The vacuum case}

Let us first review the application of the axis relation within the
context of vacuum spacetimes, where we know that the quadruple-Neugebauer
(double-Harrison) transformation does the job.  It will suffice to
consider the generation of the Kerr metric, for which
\begin{equation}
\E = 1-\frac{2m}{r-ia\cos\theta} \; .
\end{equation}
On the axis, where $\theta=0$, we have $\cos\theta=1$.  On the other hand,
the Weyl canonical coordinates $z,\rho$ are given by
\begin{equation}
\rho^{2} = (r^{2}+a^{2}-2mr)\sin^{2}\theta \; , \;
z = (r-m) \cos\theta \; .
\end{equation}
Therefore, the axis data for the Kerr metric assumes the form
\begin{equation}
\E(z,0) = \frac{z-m-ia}{z+m-ia} \; ,
\end{equation}
i.e., $U(z,0)=z-ia$ and $W(z,0)=m$.

Now, the axis relation says\cite{RetzH}
\begin{equation}
-i\E(\tau,0) = \frac{-i\E^{(0)}(\tau,0)v^{UL}(\tau) + v^{UR}(\tau)}
{-i\E^{(0)}(\tau,0)v^{LL}(\tau) + v^{LR}(\tau)} \; ,
\end{equation}
where $\E^{(0)}$ is the complex potential of the seed metric.  For
Minkowski space, $\E^{(0)}=1$.  Plugging in $\E(\tau,0)$ for the Kerr
metric, we are tempted to try
$$
v(\tau) = \left( \begin{array}{cc}
\tau - m & - a \\ a & \tau + m
\end{array} \right) \; .
$$
However, $v(\tau)$ should be an matrix of the group $SU(1,1)=SL(2,R)$.
Since the determinant has the value
$$
\det v(\tau) = z^{2} - m^{2} + a^{2} \; ,
$$
we should divide by the square root of this determinant, and select
\begin{equation}
v(\tau) = \frac{1}{\sqrt{z^{2}-m^{2}+a^{2}}}\left( \begin{array}{cc}
\tau - m & - a \\ a & \tau + m
\end{array} \right) \; .
\end{equation}

Suppose, for the moment, that $a^{2}<m^{2}$.  Then, using the identities
\begin{equation}
\tau = \frac{1}{2} \left[ (\tau + \sqrt{m^{2}-a^{2}})
+ (\tau - \sqrt{m^{2}-a^{2}}) \right] \; , \;
1 = \frac{1}{2} \left[ (\tau + \sqrt{m^{2}-a^{2}})
- (\tau - \sqrt{m^{2}-a^{2}}) \right] \; ,
\end{equation}
we can cast our expression for $v(\tau)$ into the form
\begin{equation}
v(\tau) =
\frac{1}{2}(I+J) \sqrt{\frac{\tau+\sqrt{m^{2}-a^{2}}}
{\tau-\sqrt{m^{2}-a^{2}}}}
+ \frac{1}{2}(I-J) \sqrt{\frac{\tau-\sqrt{m^{2}-a^{2}}}
{\tau+\sqrt{m^{2}-a^{2}}}} \; ,
\end{equation}
where the matrix
\begin{equation}
J := \frac{1}{\sqrt{m^{2}-a^{2}}} \left( \begin{array}{cc}
- m & - a \\ a & m
\end{array} \right)
\end{equation}
satisfies
\begin{equation}
\tr{J} = 0 \; , \; J^{2} = I \; .
\end{equation}
This result can also be expressed in the exponential form
\begin{equation}
v(\tau) = \exp(J\eta(\tau)) \; ,
\end{equation}
where
\begin{equation}
e^{\eta(\tau)} = \sqrt{\frac{\tau+\sqrt{m^{2}-a^{2}}}
{\tau-\sqrt{m^{2}-a^{2}}}} \; .
\end{equation}
If, on the other hand, $a^{2}>m^{2}$, this procedure would have
yielded
\begin{equation}
v(\tau) =
\frac{1}{2}(I+J) \sqrt{\frac{\tau+i\sqrt{a^{2}-m^{2}}}
{\tau-i\sqrt{a^{2}-m^{2}}}}
+ \frac{1}{2}(I-J) \sqrt{\frac{\tau-i\sqrt{a^{2}-m^{2}}}
{\tau+i\sqrt{a^{2}-m^{2}}}} \; ,
\end{equation}
where
\begin{equation}
iJ := \frac{1}{\sqrt{a^{2}-m^{2}}} \left( \begin{array}{cc}
- m & - a \\ a & m
\end{array} \right) \; .
\end{equation}
In either case, $v(\tau)$ is an $SU(1,1)=SL(2,R)$ matrix.  In
the case $a^{2}>m^{2}$ we probably would have used the symbol
$J$ for the real matrix $iJ$ and then would have had $J^{2}=-1$
for that case.

Alternatively, we can unify these two cases by considering members
of the larger group $SL(2,C)$, temporarily setting aside the reality
condition on $v(\tau)$ and allowing general complex values for the
parameters.  It turns out\cite{RetzH} that the Hauser-Ernst homogeneous
Hilbert problem works for members of $SL(2,C)$ as well as $SU(1,1)=SL(2,R)$,
so there is no need to solve the HHP twice.  The solutions for both
$a^{2}<m^{2}$ and $a^{2}>m^{2}$ can be inferred from the complexified
spacetime that results from an application of the $SL(2,C)$
transformation.

\subsection{The electrovac case}

Now let us turn our attention to the charged Kerr metric, where
\begin{equation}
\E = 1 - \frac{2m}{r-ia\cos\theta} \; , \;
\Phi = \frac{e}{r-ia\cos\theta} \; .
\end{equation}
Thus,
\begin{equation}
\E(\tau,0) = \frac{\tau-m-ia}{\tau+m-ia} \; , \;
\Phi(\tau,0) = \frac{e}{\tau+m-ia} \; .
\end{equation}
In the electrovac case the axis relation can be expressed in the
form\cite{RetzE}
\begin{equation}
X(\tau) v(\tau) Y(\tau) = 0 \; , \;
X(\tau) v(\tau) Z(\tau) = 0 \; ,
\end{equation}
where
\begin{equation}
X(\tau) := \left( \begin{array}{ccc}
-\frac{1}{\sqrt{2}} i & \frac{1}{\sqrt{2}} \E(\tau,0) & \Phi(\tau,0)
\end{array} \right) \; ,
\end{equation}
and
\begin{equation}
Y(\tau) := \left( \begin{array}{c}
- i \E^{(0)}(\tau,0) \\ 1 \\ 0
\end{array} \right) \; , \;
Z(\tau) := \left( \begin{array}{c}
- \frac{1}{\sqrt{2}} i \Phi^{(0)}(\tau,0) \\ 0 \\ 1
\end{array} \right) \; .
\end{equation}
In the present case
\begin{equation}
X(\tau) = \left( \begin{array}{ccc}
- \frac{1}{\sqrt{2}} i (\tau+m-ia) & \frac{1}{\sqrt{2}} (\tau-m-ia) & e
\end{array} \right) \; ,
\end{equation}
and
\begin{equation}
Y(\tau) = \left( \begin{array}{c}
- i \\ 1 \\ 0
\end{array} \right) \; , \;
Z(\tau) = \left( \begin{array}{c}
0 \\ 0 \\ 1
\end{array} \right) \; .
\end{equation}
The resulting conditions upon the $SU(2,1)$ matrix $v(\tau)$ are not
as simple as in the vacuum case, but there is a certain resemblance
nevertheless.  Guided by how the HHP was solved in the vacuum case,
we look for a $v(\tau)$ of the form
\begin{equation}
v(\tau) = \exp(J\eta(\tau)) \; ,
\end{equation}
where two of the three eigenvalues of $J$ are degenerate.  Because
$\tr{J}=0$, these two eigenvalues are $\lambda$ and $-2\lambda$,
respectively, where $\lambda$ is a constant.  If $\P$ is a projection
operator onto the subspace corresponding to the nondegenerate
eigenvector, we may express $v(\tau)$ in the form
\begin{equation}
v(\tau) = (I-\P) e^{\lambda\eta(\tau)}
+ \P e^{-2\lambda\eta(\tau)} \; ,
\end{equation}
where the projection operator $\P$ can be written in term of the
nondegenerate eigenvector $h$ of $J$ as
\begin{equation}
\P = (1/E) h h^{\dagger} \Omega \; , \; E := h^{\dagger} \Omega h \; ,
\end{equation}
where
\begin{equation}
\Omega := \left( \begin{array}{ccc}
0 & i & 0 \\ -i & 0 & 0 \\ 0 & 0 & 1
\end{array} \right) \; .
\end{equation}
This way of writing $\P$ implies that
\begin{equation}
\P^{\dagger} \Omega = \Omega \P \; .
\end{equation}
On the other hand,
\begin{equation}
J = \lambda (I - 3 \P) \; ,
\end{equation}
so
\begin{equation}
\lambda J^{\dagger}\Omega - \lambda^{*} \Omega J = 0 \; .
\end{equation}
On the other hand, the $SU(2,1)$ conditions
\begin{equation}
\det{v} = 1 \; , \; v^{\dagger} \Omega v = \Omega \; ,
\end{equation}
require
\begin{equation}
\eta(\tau)^{*} J^{\dagger} \Omega + \eta(\tau) \Omega J = 0 \; ,
\end{equation}
which is satisfied if and only if $\lambda\eta(\tau)$ is imaginary.

We may now express the axis relations in the form
\begin{mathletters}
\begin{eqnarray}
X(\tau) \left[ I + \P \left( e^{-3\lambda\eta(\tau)} - 1 \right) \right]
Y(\tau) & = & 0 \; , \\
X(\tau) \left[ I + \P \left( e^{-3\lambda\eta(\tau)} - 1 \right) \right]
Z(\tau) & = & 0 \; .
\end{eqnarray}
\end{mathletters}
However, in our case
\begin{equation}
X(\tau) Y(\tau) = - \sqrt{2} m \text{ and } X(\tau) Z(\tau) = e
\end{equation}
are both constants.  Thus, our pair of equations reduces to
\begin{mathletters}
\begin{eqnarray}
\left( e^{-3\lambda\eta(\tau)} - 1 \right) X(\tau) h h^{\dagger} \Omega
Y & = & \sqrt{2} E m \; , \\
\left( e^{-3\lambda\eta(\tau)} - 1 \right) X(\tau) h h^{\dagger} \Omega
Z & = & - E e \; .
\end{eqnarray}
\end{mathletters}
Because
\begin{equation}
h^{\dagger} \Omega Y = i (h_{1}^{*} + i h_{2}^{*}) \; , \;
h^{\dagger} \Omega Z = h_{3}^{*} \; ,
\end{equation}
it is immediately apparent that
\begin{equation}
\Q := \sqrt{2} ih_{3}/(h_{1}^{*}+ih_{2}^{*}) = e/m \; .
\end{equation}
We also find that
\begin{equation}
X(\tau) h = - i \left\{ \left( \frac{h_{1}+ih_{2}}{\sqrt{2}} \right)
(\tau - ia) + \left( \frac{h_{1}-ih_{2}}{\sqrt{2}} \right) m
+ ih_{3} e \right\} \; .
\end{equation}
Of course, we do not expect the eigenvector to be determined completely.
Rather, only the ratios of its components will be determined.  Therefore,
we shall impose an additional condition; for example,
\begin{equation}
\frac{h_{1}+ih_{2}}{\sqrt{2}} = 1 \; .
\end{equation}
Thus, we end up with the simple result
\begin{equation}
X(\tau) h = - i \left\{ (\tau - ia) + \frac{ih_{3}}{e} (e^{2}-m^{2})
\right\} \; .
\end{equation}
Moreover,
\begin{equation}
E = 1 + (|h_{3}|/e)^{2} (e^{2} - m^{2}) \; ,
\end{equation}
so the axis relation yields
\begin{equation}
\left( e^{-3\lambda\eta(\tau)} - 1 \right)
\left[ \tau - ia + \frac{ih_{3}}{e} (e^{2} - m^{2}) \right]
= - \left[ \frac{ie}{h_{3}^{*}} + \frac{ih_{3}}{e} (e^{2} - m^{2}) \right] \; ,
\end{equation}
or
\begin{equation}
e^{3\lambda\eta(\tau)} = \frac{\tau-ia+i(h_{3}/e)(e^{2}-m^{2})}
{\tau-ia-i(e/h_{3}^{*})} \; . \label{tentative}
\end{equation}

To complete the discussion we must fully determine the eigenvector $h$,
and hence the projection operator $\P$.  Only $h_{3}$ remains to be
determined, for once $h_{3}$ is known, we shall also know
$(h_{1}-ih_{2})/\sqrt{2}$, and $(h_{1}+ih_{2})/\sqrt{2}=1$ by our
convention for the selection of the representative eigenvector $h$.
Suppose now that $e/h_{3}$ is a root of the equation
\begin{equation}
(e/h_{3})^{2} + 2a (e/h_{3}) - (e^{2}-m^{2}) = 0 \; .
\end{equation}
If $a^{2}+e^{2}>m^{2}$, there are two real roots
\begin{equation}
e/h_{3} = - a \pm \sqrt{a^{2}+e^{2}-m^{2}} \; ,
\end{equation}
and, if $a^{2}+e^{2}<m^{2}$, there are two complex conjugate roots
\begin{equation}
e/h_{3} = - a \pm i \sqrt{m^{2}-a^{2}-e^{2}} \; .
\end{equation}
Finally, substituting the first of these roots back into Eq.\
(\ref{tentative}), we obtain
\begin{equation}
e^{3\lambda\eta(\tau)} = \frac{\tau+i\sqrt{a^{2}+e^{2}-m^{2}}}
{\tau-i\sqrt{a^{2}+e^{2}-m^{2}}} \; ,
\end{equation}
and substituting the second of these roots into Eq.\ (\ref{tentative}),
we obtain
\begin{equation}
e^{3\lambda\eta(\tau)} = \frac{\tau+\sqrt{m^{2}-a^{2}-e^{2}}}
{\tau-\sqrt{m^{2}-a^{2}-e^{2}}} \; .
\end{equation}
In the first case, $\lambda\eta(\tau)$ is imaginary, while in the
second case, $\lambda\eta(\tau)$ is real.  Thus, only in the first
case is the $SU(2,1)$ condition satisfied by $v(\tau)$.

One may infer from this disappointing result that an $SU(2,1)$
matrix $v(\tau)$ for the case $a^{2}+e^{2}<m^{2}$ must {\em not} have
two degenerate eigenvectors.  That complicates the identification of
a suitable transformation.  As far as we know, no one has yet worked out
an $SU(2,1)$ transformation matrix $v(\tau)$ that accomplishes our
purposes when $a^{2}+e^{2}<m^{2}$, let alone solved the associated HHP.
On the other hand, the $SU(2,1)$ transformation matrix $v(\tau)$ that we
found for the case $a^{2}+e^{2}>m^{2}$ corresponds to a transformation
that was discovered by Alekseev\cite{Alekseev} and by
Cosgrove\cite{Cosgrove} many years ago.

Like the double-Harrison transformation, the Cosgrove transformation
can be complexified, i.e., the $SU(2,1)$ transformation matrix $v(\tau)$
can be replaced by an $SL(3,C)$ matrix.  The solution corresponding to
$a^{2}+e^{2}>m^{2}$ will then correspond to an obvious {\em real cross
section} of the complexified spacetime that results from the application
of the complexified Cosgrove transformation to Minkowski space.  The big
question is, however, ``Can one identify another real cross section that
corresponds to the case $a^{2}+e^{2}<m^{2}$?

In the case of the charged Kerr metric it is fairly trivial to infer
the $\E$ and $\Phi$ potentials and the metric fields for the case
$a^{2}+e^{2}<m^{2}$ from the corresponding potentials and fields for
the case $a^{2}+e^{2}>m^{2}$.  The author has always believed that
this type of construction would be possible for all electrovac spacetimes
that belong to the Cosgrove family, but only recently, after recognizing
the formal similarity of a five-parameter electrovac solution obtained by
Manko et al.\cite{Manko} to a particular specialization of a
twelve-parameter solution that was generated many years ago by Guo
and Ernst\cite{GE} using the Cosgrove transformation, has he actually
tried to prove that this is indeed possible.

\section{Complexified Cosgrove Transformation}

The form of $U$, $V$ and $W$ that we shall present for the spacetime that
results from applying a succession of $n$ Cosgrove transformations of
Minkowski space is new, and was derived in the following way from
expressions the reader can find in earlier work of Cosgrove,\cite{Cosgrove}
Guo and Ernst,\cite{GE} Chen, Guo and Ernst\cite{CGE} and Wang, Guo and
Wu.\cite{WGW}

For the first Cosgrove transformation, Guo and Ernst expressed the
complex potentials in the form\cite{change}
\begin{equation}
\E = 1 - 2i \frac{N}{D} \; , \; \Phi = - \frac{N'}{D} \; ,
\end{equation}
where $D$ was written as the determinant of a $3 \times 3$ matrix, the
columns of which were denoted by $\psi_{1}$, $\psi_{2}$ and $\psi_{3}$,
and which were proportional, respectively, to
\begin{equation}
P^{(0)}(K^{*})h \; , \; P^{(0)}(K)h' \; \text{ and } P^{(0)}(K)h'' \; ,
\end{equation}
where
\begin{equation}
P^{(0)}(\tau) := \frac{1}{\sqrt{2}r(\tau)} \left( \begin{array}{ccc}
-[r(\tau)-(\tau-z)] & i[r(\tau)+(\tau-z)] & 0 \\
-i &  1 & 0 \\
0 & 0 & \sqrt{2}r(\tau)
\end{array} \right)
\end{equation}
is the $P$-potential\cite{RetzE} of Minkowski space,
\begin{equation}
r(\tau) := \sqrt{(z-\tau)^{2}+\rho^{2}} \; ,
\end{equation}
and $K$ is a complex parameter.  The elements of the vectors $h$
are arbitrary complex parameters, but only ratios of these components
are significant.  The vectors $h'$ and $h''$ are linearly independent
vectors that are ``orthogonal'' to $h$ in the sense
\begin{equation}
h^{\dagger} \Omega h' = 0
= h^{\dagger} \Omega h'' \; ,
\end{equation}
where
\begin{equation}
\Omega := \left( \begin{array}{ccc}
0 & i & 0 \\ -i & 0 & 0 \\ 0 & 0 & 1
\end{array} \right) \; .
\end{equation}
Explicit expressions for the components of the vectors $\psi^{(k)}_{i}$
were given by Chen, Guo and Ernst in Eqs.~(8).  However, to obtain
simpler expressions, we shall select $h''$ differently than they did.
If one chooses the vectors
\begin{equation}
h = \left( \begin{array}{c}
h_{1} \\ h_{2} \\ h_{3}
\end{array} \right) \; , \;
h' = \left( \begin{array}{c}
h^{*}_{1} \\ h^{*}_{2} \\ 0
\end{array} \right) \; , \;
h'' = \left( \begin{array}{c}
h^{*}_{3} \\ ih^{*}_{3} \\ h^{*}_{1} + ih^{*}_{2}
\end{array} \right) \; ,
\end{equation}
then one can select
\begin{equation}
\psi_{1} = \left( \begin{array}{c}
Q_{1} \\ 1 \\ \Q^{*} S_{1}
\end{array} \right) \; , \;
\psi_{2} = \left( \begin{array}{c}
Q_{2} \\ 1 \\ 0
\end{array} \right) \; , \;
\psi_{3} = \left( \begin{array}{c}
i\Q \\ 0 \\ 1
\end{array} \right) \; , \label{psi}
\end{equation}
where $K_{1} := K^{*}$, $K_{2} := K$,
\begin{equation}
Q_{k} := i [ X_{k}r_{k} + (K_{k} - z) ] \; , \;
S_{k} := X_{k}r_{k} \; , \label{Q}
\end{equation}
and
\begin{equation}
X_{1} := - \frac{h_{1}-ih_{2}}{h_{1}+ih_{2}} \; , \;
X_{2} := - \frac{h_{1}^{*}-ih_{2}^{*}}{h_{1}^{*}+ih_{2}^{*}} \; , \;
\Q := \sqrt{2} i \frac{h_{3}^{*}}{h_{1}^{*}+ih_{2}^{*}} \; .
\end{equation}
We note, in particular, that
\begin{equation}
K_{2} = K_{1}^{*} \text{ and } X_{1}^{*} X_{2} = 1 \; . \label{restrict}
\end{equation}

Guo and Ernst also constructed the complex potentials $\E$ and $\Phi$
for the spacetime that results when Minkowski space is subjected to two
successive Cosgrove transformations.  Wang, Guo and Wu then showed that
$D$, $N$ and $N'$ could be reexpressed as $6 \times 6$ determinants,
while, for higher values of $n$, they could be expressed as
$3n \times 3n$ determinants.  In our present gauge the Wang-Guo-Wu
expression for $D$ assumes the simple form
\begin{equation}
D = \left| \begin{array}{ccc}
D_{11} & \cdots & D_{1n} \\
\vdots & & \vdots \\
D_{n1} & \cdots & D_{nn}
\end{array} \right| \; ,
\end{equation}
where the $3 \times 3$ submatrices $D_{jk}$ are given by
\begin{equation}
D_{jk} := \left( \begin{array}{ccc}
(K_{2k-1})^{j-1}Q_{2k-1} & (K_{2k})^{j-1}Q_{2k}
	& i(K_{2k})^{j-1}\Q_{k} \\
(K_{2k-1})^{j-1} & (K_{2k})^{j-1} & 0 \\
(K_{2k-1})^{j-1}\Q_{k}^{*}S_{2k-1} & 0 & (K_{2k})^{j-1}
\end{array} \right) \; .
\end{equation}
The determinants $N$ and $N'$ can be constructed from $D$ by replacing,
respectively, the $(3n)$-th and $(3n-2)$-nd rows by
$$
K_{1}^{n} \quad K_{2}^{n} \quad 0 \quad
	\cdots \quad K_{2n-1}^{n} \quad K_{2n}^{n} \quad 0 \; .
$$

The fields $U$, $V$ and $W$ are defined (up to a common factor) by
\begin{equation}
U := D-iN \; , \; V := -N' \; \text{ and } W := iN \; ,
\end{equation}
each of which can obviously be written as a single determinant.  By
using elementary row operations it can be shown that the term $i(K_{k}-z)$
in $Q_{k} (k=1,\ldots,2n)$ contributes nothing to any of the determinants.
Therefore, each $Q_{k}$ can be replaced by $iS_{k}$.  In conclusion,
the complex potentials $\E$ and $\Phi$ of the electrovac solution that
results from applying a succession of $n$ Cosgrove transformations to
Minkowski space are given by
\begin{equation}
\E = \frac{U-W}{U+W} \; , \; \Phi = \frac{V}{U+W} \; ,
\end{equation}
where $U$ is the $3n \times 3n$ determinant
\begin{equation}
U = \left| \begin{array}{ccc}
U_{11} & \cdots & U_{1n} \\
\vdots & & \vdots \\
U_{n1} & \cdots & U_{nn}
\end{array} \right| \; ,
\end{equation}
in which occur the $3 \times 3$ submatrices
\begin{equation}
U_{jk} := \left( \begin{array}{ccc}
(K_{2k-1})^{j-1}X_{2k-1}r_{2k-1} & (K_{2k})^{j-1}X_{2k}r_{2k}
	& (K_{2k})^{j-1}\Q_{k} \\
(K_{2k-1})^{j-1} & (K_{2k})^{j-1} & 0 \\
(K_{2k-1})^{j-1}\Q_{k}^{*}X_{2k-1}r_{2k-1} & 0 & (K_{2k})^{j-1}
\end{array} \right) \; ,
\end{equation}
where
\begin{equation}
r_{a} := \sqrt{(z-K_{a})^{2}+\rho^{2}} \; .
\end{equation}
The $3n \times 3n$ determinants $-V$ and $W$ are constructed from $U$ by
replacing, respectively, the $(3n)$-th and the $(3n-2)$-nd row of the
latter determinant by
$$
(K_{1})^{n} \quad (K_{2})^{n} \quad 0 \quad
	\cdots \quad (K_{2n-1})^{n} \quad (K_{2n})^{n} \quad 0 \; .
$$

We have seen that the Cosgrove transformation, which is characterized by
a $v(\tau)$ with one non-degenerate eigenvector $h$, and a pair of
degenerate eigenvectors $h'$ and $h''$, cannot cover all cases.  A
Kinnersley-Chitre transformation $v(\tau)$ with three non-degenerate
eigenvectors is needed as well, but such a transformation has not yet
been identified, and the associated HHP has not yet been solved.

By the {\em complexified Cosgrove transformation} we shall mean the
$SL(3,C)$ matrix $v(\tau)$ with eigenvectors such that Eqs.\ (\ref{psi})
through (\ref{restrict}) are valid, but $X_{1}^{*}$, $X_{2}^{*}$ and
$\Q^{*}$ are complex parameters that are no longer identified as the
complex conjugates of $X_{1}$, $X_{2}$ and $\Q$, respectively.  Similarly,
$K_{1}^{*}=K_{2}$ is no longer the complex conjugate of $K_{1}$, and
$K_{2}^{*}=K_{1}$ is no longer the complex conjugate of $K_{2}$.  The
$n$-fold complexified Cosgrove transformation is characterized by
$6n$ complex parameters, while the ordinary Cosgrove transformation is
characterized by only $3n$ complex parameters.  The complexified
Cosgrove transformation for $n=2$ has sufficiently many complex
parameters to cover all axis data of the form
\begin{mathletters}
\begin{eqnarray}
U(z,0) & = & z^{2} + U_{1} z + U_{2} \; , \\
V(z,0) & = & V_{1} z + V_{2} \; , \\
W(z,0) & = & W_{1} z + W_{2} \; , \\
U^{*}(z,0) & = & z^{2} + U_{1}^{*} z + U_{2}^{*} \; , \\
V^{*}(z,0) & = & V_{1}^{*} z + V_{2}^{*} \; , \\
W^{*}(z,0) & = & W_{1}^{*} z + W_{2}^{*} \; ,
\end{eqnarray}
\end{mathletters}
where the complex fields $U^{*}$, $V^{*}$ and $W^{*}$ are regarded as
independent of the complex fields $U$, $V$ and $W$, and where
$U_{1}$, $U_{2}$, $V_{1}$, $V_{2}$, $W_{1}$, $W_{2}$, $U_{1}^{*}$,
$U_{2}^{*}$, $V_{1}^{*}$, $V_{2}^{*}$, $W_{1}^{*}$ and $W_{2}$ are
twelve independent complex constants.  These fields $U,V,W$ and
$U^{*},V^{*},W^{*}$ satisfy the field equations
\begin{mathletters}
\begin{eqnarray}
(U^{*}U+V^{*}V-W^{*}W) \; \nabla^{2} \left( \begin{array}{c}
U \\ V \\ W
\end{array} \right) & = &
2 (U^{*} \nabla U + V^{*} \nabla V - W^{*} \nabla W) \cdot
\nabla \left( \begin{array}{c}
U \\ V \\ W
\end{array} \right) \; , \\
(U^{*}U+V^{*}V-W^{*}W) \; \nabla^{2} \left( \begin{array}{c}
U^{*} \\ V^{*} \\ W^{*}
\end{array} \right) & = &
2 (U \nabla U^{*} + V \nabla V^{*} - W \nabla W^{*}) \cdot
\nabla \left( \begin{array}{c}
U^{*} \\ V^{*} \\ W^{*}
\end{array} \right) \; .
\end{eqnarray}
\end{mathletters}
One should be aware of the fact that the metric field $f$, which is
defined by
\begin{equation}
f := \frac{U^{*}U+V^{*}V-W^{*}W}{(U^{*}+W^{*})(U+W)} \; ,
\end{equation}
is, in general, complex, since $U^{*}$ is no longer the complex
conjugate of $U$, etc.  For this reason, we refer to these as
{\em complexified spacetimes}, even though the $z$ and $\rho$
coordinates remain real.

In the case of the complexified Cosgrove transformation, there are
expressions for the independent complex potentials $\E^{*}$ and
$\Phi^{*}$ that precisely parallel the expressions we have already
given for $\E$ and $\Phi$.  We shall not state these relations
explicitly, since they can be constructed quite easily by the reader.
When $\E^{*}$ is the complex conjugate of $\E$ and $\Phi^{*}$ is the
complex conjugate of $\Phi$, we shall say that we have a {\em real
cross section} of the complexified spacetime.

When the complex constants $\Q_{k}$ and $\Q^{*}_{k}$ all vanish,
we obtain a vacuum solution, the real cross sections of which are
the vacuum metrics of the Neugebauer family, the $n=2$ exemplars
of which were studied in Ref.~1.

\section{The $n=2$ Solution}

In this paper, as in Ref.~1, we shall restrict our attention to the case
$n=2$, where $U$, $V$ and $W$ are given by
\begin{mathletters}
\begin{eqnarray}
U & = &
(K_{1}-K_{4})(K_{2}-K_{3})
\left[ (1-|\Q_{1}|^{2}) X_{1}r_{1} - X_{2}r_{2} \right]
\left[ (1-|\Q_{2}|^{2}) X_{3}r_{3} - X_{4}r_{4} \right]
\nonumber \\ & & \mbox{}
+ (K_{1}-K_{2})(K_{3}-K_{4})
\left[ (1-\Q_{1}^{*}\Q_{2}) X_{1}r_{1} - X_{4}r_{4} \right]
\left[ (1-\Q_{1}\Q_{2}^{*}) X_{3}r_{3} - X_{2}r_{2} \right]
\label{U} \; , \\
V & = & \mbox{} - \Delta(K_{2},K_{3},K_{4}) Y_{1}r_{1}
+ \Delta(K_{3},K_{4},K_{1}) Y_{2}r_{2}
\nonumber \\ & & \mbox{}
- \Delta(K_{4},K_{1},K_{2}) Y_{3}r_{3}
+ \Delta(K_{1},K_{2},K_{3}) Y_{4}r_{4}
\label{V} \; , \\
W & = & \mbox{} - \Delta(K_{2},K_{3},K_{4}) Z_{1}r_{1}
+ \Delta(K_{3},K_{4},K_{1}) Z_{2}r_{2}
\nonumber \\ & & \mbox{}
- \Delta(K_{4},K_{1},K_{2}) Z_{3}r_{3}
+ \Delta(K_{1},K_{2},K_{3}) Z_{4}r_{4}
\label{W} \; ,
\end{eqnarray}
\end{mathletters}
where
\begin{mathletters}
\begin{eqnarray}
Y_{1} & := & \left[
\Q_{2} (1 - |\Q_{1}|^{2}) + \frac{K_{1}-K_{2}}{K_{4}-K_{2}}
(\Q_{1}-\Q_{2}) \right] X_{1} \; , \\
Y_{2} & := & \Q_{2} X_{2} \; , \\
Y_{3} & = & \left[
\Q_{1} (1-|\Q_{2}|^{2}) - \frac{K_{4}-K_{3}}{K_{4}-K_{2}}(\Q_{1}-\Q_{2})
\right] X_{3} \; , \\
Y_{4} & := & \Q_{1} X_{4} \; ,
\end{eqnarray}
\end{mathletters}
and
\begin{mathletters}
\begin{eqnarray}
Z_{1} & := & \left[
(1-|\Q_{1}|^{2}) + \frac{K_{1}-K_{2}}{K_{4}-K_{2}} \Q_{1}^{*} (\Q_{1}-\Q_{2})
\right] X_{1} \; , \\
Z_{2} & := & X_{2} \; , \\
Z_{3} & := & \left[
(1-|\Q_{2}|^{2}) - \frac{K_{4}-K_{3}}{K_{4}-K_{2}} \Q_{2}^{*} (\Q_{1}-\Q_{2})
\right] X_{3} \; , \\
Z_{4} & := & X_{4} \; ,
\end{eqnarray}
\end{mathletters}
and $\Delta$ is the Vandermonde determinant introduced in Ref.~1.
(We have divided out a common factor $K_{4}-K_{2}$ from $U$, $V$ and
$W$.)

{}From the above expression for $U$, one easily identifies the complex
constants
\begin{mathletters}
\begin{eqnarray}
U_{0} & = &
(K_{1}-K_{4})(K_{2}-K_{3})
\left[ (1-|\Q_{1}|^{2}) X_{1} - X_{2} \right]
\left[ (1-|\Q_{2}|^{2}) X_{3} - X_{4} \right]
\nonumber \\ & & \mbox{}
+ (K_{1}-K_{2})(K_{3}-K_{4})
\left[ (1-\Q_{1}^{*}\Q_{2}) X_{1} - X_{4} \right]
\left[ (1-\Q_{1}\Q_{2}^{*}) X_{3} - X_{2} \right] \; , \\
U_{1} & = & \mbox{}
- (K_{1}-K_{4})(K_{2}-K_{3}) \left\{
\left[ (1-|\Q_{1}|^{2}) X_{1} - X_{2} \right]
\left[ (1-|\Q_{2}|^{2}) K_{3}X_{3} - K_{4}X_{4} \right]
\right. \nonumber \\ & & \hspace{1in} \left. \mbox{}
+ \left[ (1-|\Q_{1}|^{2}) K_{1}X_{1} - K_{2}X_{2} \right]
\left[ (1-|\Q_{2}|^{2}) X_{3} - X_{4} \right]
\right\}
\nonumber \\ & & \mbox{}
- (K_{1}-K_{2})(K_{3}-K_{4}) \left\{
\left[ (1-\Q_{1}^{*}\Q_{2}) X_{1} - X_{4} \right]
\left[ (1-\Q_{1}\Q_{2}^{*}) K_{3}X_{3} - K_{2}X_{2} \right]
\right. \nonumber \\ & & \hspace{1in} \left. \mbox{}
+ \left[ (1-\Q_{1}^{*}\Q_{2}) K_{1}X_{1} - K_{4}X_{4} \right]
\left[ (1-\Q_{1}\Q_{2}^{*}) X_{3} - X_{2} \right]
\right\} \; , \\
U_{2} & = &
(K_{1}-K_{4})(K_{2}-K_{3})
\left[ (1-|\Q_{1}|^{2}) K_{1}X_{1} - K_{2}X_{2} \right]
\left[ (1-|\Q_{2}|^{2}) K_{3}X_{3} - K_{4}X_{4} \right]
\nonumber \\ & &
+ (K_{1}-K_{2})(K_{3}-K_{4})
\left[ (1-\Q_{1}^{*}\Q_{2}) K_{1}X_{1} - K_{4}X_{4} \right]
\left[ (1-\Q_{1}\Q_{2}^{*}) K_{3}X_{3} - K_{2}X_{2} \right] ,
\end{eqnarray}
\end{mathletters}
while from the expressions for $V$ and $W$ one obtains the complex
constants
\begin{equation}
V_{1} = - V^{(0)} \; , \; V_{2} = V^{(1)} \; , \;
W_{1} = - W^{(0)} \; , \; W_{2} = W^{(1)} \; ,
\end{equation}
where
\begin{mathletters}
\begin{eqnarray}
V^{(a)} & := & \mbox{}
- \Delta(K_{2},K_{3},K_{4})K_{1}^{a}Y_{1}
+ \Delta(K_{3},K_{4},K_{1})K_{2}^{a}Y_{2}
\nonumber \\ & & \mbox{}
- \Delta(K_{4},K_{1},K_{2})K_{3}^{a}Y_{3}
+ \Delta(K_{1},K_{2},K_{3})K_{4}^{a}Y_{4} \; , \\
W^{(a)} & := & \mbox{}
- \Delta(K_{2},K_{3},K_{4})K_{1}^{a}Z_{1}
+ \Delta(K_{3},K_{4},K_{1})K_{2}^{a}Z_{2}
\nonumber \\ & & \mbox{}
- \Delta(K_{4},K_{1},K_{2})K_{3}^{a}Z_{3}
+ \Delta(K_{1},K_{2},K_{3})K_{4}^{a}Z_{4} \; .
\end{eqnarray}
\end{mathletters}
Our immediate objective is to determine the parameters $X_{a}$, $Y_{a}$,
and $Z_{a} \; (a=1,2,3,4)$ in terms of the axis data $U_{1}$, $U_{2}$,
$V_{1}$, $V_{2}$, $W_{1}$ and $W_{2}$ (where $U_{0} = 1$), and the
$K$'s.

\subsection{Determination of $X_{a} \; (a=1,2,3,4)$}

The simplest case is when $V_{2}W_{1}-V_{1}W_{2}=0$.  This corresponds
to $\Q_{1}=\Q_{2}=:\Q$, where $V_{1}=\Q W_{1}$ and $V_{2}=\Q W_{2}$.
In this case one gets the same expressions for $(1-|\Q|^{2})X_{1}$,
$X_{2}$, $(1-|\Q|^{2})X_{3}$ and $X_{4}$ as one got for $X_{1}$,
$X_{2}$, $X_{3}$ and $X_{4}$, respectively, in the vacuum case.  This
solution is merely the electrovac solution that is generated by the
old electrification transformation of Harrison, as reformulated by
Ernst.  In this paper, we are concerned primarily with the case when
$V_{2}W_{1}-V_{1}W_{2} \ne 0$, i.e., $\Q_{1} \ne \Q_{2}$.

In the vacuum problem it was surprisingly easy to solve for the $X$'s
in terms of the axis data and the $K$'s.  One had to solve nothing but
linear and quadratic algebraic equations.  The solution itself revealed
a most intriguing structure, to which the simplicity of the solution can
be attributed, and a knowledge of which we found to be indispensable for
solving the electrovac problem.  On the one hand, one had the hierarchy
of linear equations (for the $X$'s)
\begin{mathletters}
\begin{eqnarray}
W^{(0)} & = & - W_{1} \; , \\
W^{(1)} & = & W_{2} \; , \\
W^{(2)} & = &
- \frac{U_{1} W_{2} - U_{2} W_{1}}{U_{0}} \; , \\
W^{(3)} & = &
\frac{U_{1}\left[(U_{1} W_{2} - U_{2} W_{1})/U_{0}\right]
- U_{2} W_{2}}{U_{0}} \; ,
\end{eqnarray}
\end{mathletters}
and, on the other hand,
\begin{equation}
\Delta(K_{1},K_{2},K_{3},K_{4}) =
-\frac{W_{1}\left[(U_{1}W_{2}-U_{2}W_{1})/U_{0}\right]
-W_{2}^{2}}{U_{0}} \; .
\end{equation}
In particular, the last equation allowed one to determine $U_{0}$
in terms of the axis data and the $K$'s, a critical step in the
complete determination of the $X$'s.

After spending a considerable amount of time trying to identify the
$X$'s and the $\Q$'s in the electrovac case, we abandoned that effort,
and approached the problem of determining $U$, $V$ and $W$ in a new
way that avoids the determination of the $X$'s and $\Q$'s (although
these objects can be calculated at the very end, if they are really
desired).  In the electrovac case we have two sets of four linear
equations,
\begin{mathletters}
\begin{eqnarray}
V^{(0)} & = & -V_{1} \; , \label{V0} \\
V^{(1)} & = & V_{2} \; , \\
V^{(2)} & = & \frac{U_{2}V_{1}-U_{1}V_{2}}{U_{0}} \; , \\
V^{(3)} & = & -\frac{U_{2}V_{2}+U_{1}[(U_{2}V_{1}-U_{1}V_{2})/U_{0}]}
{U_{0}} + \frac{V_{2}W_{1}-V_{1}W_{2}}{U_{0}}
\left(\frac{W_{1}}{U_{0}}\right)^{*} \; , \label{V3}
\end{eqnarray}
\end{mathletters}
and
\begin{mathletters}
\begin{eqnarray}
W^{(0)} & = & -W_{1} \; , \label{W0} \\
W^{(1)} & = & W_{2} \; , \\
W^{(2)} & = & \frac{U_{2}W_{1}-U_{1}W_{2}}{U_{0}} \; , \\
W^{(3)} & = & -\frac{U_{2}W_{2}+U_{1}[(U_{2}W_{1}-U_{1}W_{2})/U_{0}]}
{U_{0}} + \frac{V_{2}W_{1}-V_{1}W_{2}}{U_{0}}
\left(\frac{V_{1}}{U_{0}}\right)^{*} \; , \label{W3}
\end{eqnarray}
\end{mathletters}
As in the vacuum case, the right side of each of the four equations is
equal to $U_{0}$ times a quantity that can be easily expressed in terms
of the axis data alone, while the left side of each of the four
equations is equal to a linear combination of $Y_{a} \; (a=1,2,3,4)$
or $Z_{a} \; (a=1,2,3,4)$.

\subsection{Determination of $U$, $W$ and $V$ up to a common factor}

The four linear equations (\ref{V0})---(\ref{V3}) for $Y_{a} \;
(a=1,2,3,4)$ and the four linear equations (\ref{W0})---(\ref{W3})
for $Z_{a} \; (a=1,2,3,4)$ are easily solved.  One obtains
\begin{eqnarray}
\D Y_{1} & = & \{U_{2}V_{2}
+U_{1}(U_{2}V_{1}-U_{1}V_{2}) - (V_{2}W_{1}-V_{1}W_{2}) W_{1}^{*} \}
\nonumber \\ & & \mbox{}
+ (K_{2}+K_{3}+K_{4}) (U_{2}V_{1}-U_{1}V_{2})
\nonumber \\ & & \mbox{}
- (K_{2}K_{3}+K_{2}K_{4}+K_{3}K_{4}) V_{2}
- (K_{2}K_{3}K_{4}) V_{1} \; , \label{DY1}
\end{eqnarray}
and
\begin{eqnarray}
\D Z_{1} & = & \{U_{2}W_{2}
+U_{1}(U_{2}W_{1}-U_{1}W_{2}) - (V_{2}W_{1}-V_{1}W_{2}) V_{1}^{*} \}
\nonumber \\ & & \mbox{}
+ (K_{2}+K_{3}+K_{4}) (U_{2}W_{1}-U_{1}W_{2})
\nonumber \\ & & \mbox{}
- (K_{2}K_{3}+K_{2}K_{4}+K_{3}K_{4}) W_{2}
- (K_{2}K_{3}K_{4}) W_{1} \; , \label{DZ1}
\end{eqnarray}
where $U$, $V$ and $W$ have been adjusted so that $U_{0}=1$.  The
expressions for $Y_{2}$, $Y_{3}$ and $Y_{4}$ can be inferred from
the expression for $Y_{1}$ and the expressions for $Z_{2}$, $Z_{3}$
and $Z_{4}$ can be inferred from the expression for $Z_{1}$ by
permuting indices on the $K$'s.  $\D$ is given by
\begin{equation}
\D := -\frac{\Delta(K_{1},K_{2},K_{3},K_{4})}{U_{0}} \; ,
\end{equation}
where $U_{0}$ is the {\em original} value of $U_{0}$, not $1$.

Using Eqs.\ (\ref{V}) and (\ref{W}), these expressions for $\D Y_{a},
\D Z_{a} \; (a=1,2,3,4)$ permit us to evaluate $\D V$ and $\D W$
without further ado, for we have
\begin{mathletters}
\begin{eqnarray}
\D V & = & \mbox{} - \Delta(K_{2},K_{3},K_{4}) \D Y_{1}r_{1}
+ \Delta(K_{3},K_{4},K_{1}) \D Y_{2}r_{2}
\nonumber \\ & & \mbox{}
- \Delta(K_{4},K_{1},K_{2}) \D Y_{3}r_{3}
+ \Delta(K_{1},K_{2},K_{3}) \D Y_{4}r_{4} \label{DV} \; , \\
\D W & = & \mbox{} - \Delta(K_{2},K_{3},K_{4}) \D Z_{1}r_{1}
+ \Delta(K_{3},K_{4},K_{1}) \D Z_{2}r_{2}
\nonumber \\ & & \mbox{}
- \Delta(K_{4},K_{1},K_{2}) \D Z_{3}r_{3}
+ \Delta(K_{1},K_{2},K_{3}) \D Z_{4}r_{4} \label{DW} \; ,
\end{eqnarray}
\end{mathletters}
but what about $U$, which is given by Eq.\ (\ref{U})?  Interestingly,
$U$ can be expressed directly in terms of the $Y$'s and $Z$'s as
follows:
\begin{equation}
U = -\frac{1}{2} \left( \frac{K_{4}-K_{2}}{Q_{1}-Q_{2}} \right)
\sum_{i,j,k,l} \epsilon_{ijkl} (K_{i}-K_{j}) Z_{k} Y_{l} r_{k} r_{l}
\; ,
\end{equation}
where $\epsilon_{ijkl}$ is Levi-Civita's permutation symbol.  On the
other hand,
\begin{equation}
\frac{K_{4}-K_{2}}{\Q_{1}-\Q_{2}} = \mbox{}
-\frac{\Delta(K_{1},K_{2},K_{3},K_{4}) U_{0}}{V_{2}W_{1}-V_{1}W_{2}}
= \D \left( \frac{U_{0}^{2}}{V_{2}W_{1}-V_{1}W_{2}} \right) \; .
\end{equation}
Hence, with $U_{0}=1$, $\D U$ can be expressed in the final form
\begin{equation}
\D U = -\frac{1}{2(V_{2}W_{1}-V_{1}W_{2})}
\sum_{i,j,k,l} \epsilon_{ijkl} (K_{i}-K_{j}) (\D Z_{k}) (\D Y_{l})
r_{k} r_{l} \; . \label{DU}
\end{equation}
This is most remarkable, since it means that $\D U$ involves only
the axis data, the known $\D Y$'s and $\D Z$'s and the $K$'s.
Since one is only interested in ratios of $U$, $V$ and $W$, it
suffices to know $\D U$, $\D V$ and $\D W$.  One does not have to
evaluate $\D$ itself!

\subsection{Determination of $K_{a} \; (a=1,2,3,4)$}

Using Eqs.\ (\ref{restrict}) and defining $|Z|^{2} := Z^{*}Z$
even when $Z^{*}$ is not just the complex conjugate of $Z$, we
find that, on the symmetry axis,
\begin{equation}
|U(z,0)|^{2} + |V(z,0)|^{2} - |W(z,0)|^{2}
= |U_{0}|^{2} (K_{1}-z)(K_{2}-z)(K_{3}-z)(K_{4}-z) \; ,
\end{equation}
from which it follows that each $K_{a} \; (a=1,2,3,4)$ satisfies the
quartic equation
\begin{eqnarray}
0 & = & K_{a}^{4} + 2(\Re{U_{1}}) K_{a}^{3}
+ (|U_{1}|^{2} + |V_{1}|^{2} - |W_{1}|^{2} + 2\Re{U_{2}}) K_{a}^{2}
\nonumber \\ & & \mbox{}
+2\Re{(U_{2}U_{1}^{*} + V_{2}V_{1}^{*} - W_{2}W_{1}^{*})} K_{a}
+ (|U_{2}|^{2} + |V_{2}|^{2} - |W_{2}|^{2}) \; ,
\end{eqnarray}
where $\Re{Z} := (Z+Z^{*})/2$ even when $Z^{*}$ is not just the
complex conjugate of $Z$.

Assuming that $z$ has been chosen so that $\Re{U_{1}}=0$ (or,
equivalently, $K_{1}+K_{2}+K_{3}+K_{4} = 0$), we have
\begin{equation}
K_{a}^{4} - A K_{a}^{2} - B K_{a} + C = 0 \; ,
\end{equation}
where
\begin{mathletters}
\begin{eqnarray}
A & = & |W_{1}|^{2} - |V_{1}|^{2} - |U_{1}|^{2} - 2\Re{U_{2}} \; , \\
B & = & - 2\Re{(U_{2}U_{1}^{*} + V_{2}V_{1}^{*} - W_{2}W_{1}^{*})} \; , \\
C & = & |U_{2}|^{2} + |V_{2}|^{2} - |W_{2}|^{2} \; .
\end{eqnarray}
\end{mathletters}
The general solution $K_{a} \; (a=1,2,3,4)$ of this quartic equation
is given by Eqs.\ (2.9a) through (2.9d) or Eqs.\ (2.14a)
through (2.14d) of Ref.~1.

In conclusion, the determination of the $K$'s is no more difficult in
the electrovac case than it was in the vacuum case.  Of course, when
$U(z,0)^{*}$, $V(z,0)^{*}$ and $W(z,0)^{*}$ are the complex conjugates
of $U(z,0)$, $V(z,0)$ and $W(z,0)$, respectively, the quartic equation,
like its vacuum analog, has solutions in which the $K$'s are real, rather
than occurring in complex conjugate pairs.  Such $K$'s cannot be used
with the ordinary Cosgrove transformation.  It is instead necessary to
employ the complexified Cosgrove transformation.

We would be the first to admit that the picture we have painted using
broad brushstrokes requires further refinement, which we hope to supply
in a future paper.  However, we shall turn now to an application that
already demonstrates the practical value of this approach.

\section{A simple but convincing application}

Suppose we select the axis data
\begin{equation}
U_{1} = -ia \; , \; U_{2} = b \; , \;
V_{1} = e \; , \; V_{2} = ic \; , \;
W_{1} = m \; , \; W_{2} = 0 \; , \label{data}
\end{equation}
where the parameters $a,b,e,c,m$ are real.  In this case, one has
\begin{equation}
U_{2}U_{1}^{*} + V_{2}V_{1}^{*} - W_{2}W_{1}^{*}
= b (-ia)^{*} + (ic) (e)^{*} - 0 (m)^{*} = i (ab+ce) \; ,
\end{equation}
so $\Re{(U_{2}U_{1}^{*}+V_{2}V_{1}^{*}-W_{2}W_{1}^{*})} = 0$.  Therefore,
as in Ref.~1, we may write
\begin{equation}
K_{1} = - K_{2} = \frac{1}{2} (\kappa_{+}+\kappa_{-}) \; , \;
K_{3} = - K_{4} = \frac{1}{2} (\kappa_{+}-\kappa_{-}) \; ,
\end{equation}
where $\kappa_{+}$ and $\kappa_{-}$ are given by
\begin{equation}
\kappa_{\pm} := \sqrt{m^{2}-a^{2}-e^{2}+2(\pm d - b)} \; , \;
d := \sqrt{b^{2} + c^{2}} \; .
\end{equation}

With the selected axis data, Eqs.\ (\ref{DY1}), (\ref{DZ1}) and
their analogs reduce to
\begin{mathletters}
\begin{eqnarray}
\D Y_{1} & = & (e/m) \D Z_{1} - \frac{1}{2} ic
\left[ (m^{2}-a^{2}-e^{2}) + \kappa_{+}\kappa_{-} + ia
(\kappa_{+}+\kappa_{-}) \right] \; , \\
\D Y_{2} & = & (e/m) \D Z_{2} - \frac{1}{2} ic
\left[ (m^{2}-a^{2}-e^{2}) + \kappa_{+}\kappa_{-} - ia
(\kappa_{+}+\kappa_{-}) \right] \; , \\
\D Y_{3} & = & (e/m) \D Z_{3} - \frac{1}{2} ic
\left[ (m^{2}-a^{2}-e^{2}) - \kappa_{+}\kappa_{-} + ia
(\kappa_{+}-\kappa_{-}) \right] \; , \\
\D Y_{4} & = & (e/m) \D Z_{4} - \frac{1}{2} ic
\left[ (m^{2}-a^{2}-e^{2}) - \kappa_{+}\kappa_{-} - ia
(\kappa_{+}-\kappa_{-}) \right] \; ,
\end{eqnarray}
\end{mathletters}
and
\begin{mathletters}
\begin{eqnarray}
\D Z_{1} & = & - m \left\{ i (ab+ce) + \frac{1}{2}
[\kappa_{+}(d+b)-\kappa_{-}(d-b)] \right\} \; , \\
\D Z_{2} & = & - m \left\{ i (ab+ce) - \frac{1}{2}
[\kappa_{+}(d+b)-\kappa_{-}(d-b)] \right\} \; , \\
\D Z_{3} & = & - m \left\{ i (ab+ce) + \frac{1}{2}
[\kappa_{+}(d+b)+\kappa_{-}(d-b)] \right\} \; , \\
\D Z_{4} & = & - m \left\{ i (ab+ce) - \frac{1}{2}
[\kappa_{+}(d+b)+\kappa_{-}(d-b)] \right\} \; ,
\end{eqnarray}
\end{mathletters}
respectively.  Eqs.\ (\ref{DU}), (\ref{V}) and (\ref{W}) then yield
the following expressions for $\D U$, $\D V$ and $\D W$:
\begin{mathletters}
\begin{eqnarray}
\D U & = &
\kappa_{-}^{2} \left\{ [ d (m^{2}-a^{2}-e^{2}) + c^{2}
- a(ab+ce) ] (r_{1}r_{3}+r_{2}r_{4})
\nonumber \right. \\ & & \hspace{0.5in} \left. \mbox{}
+ i\kappa_{+} (ab+ce+ad) (r_{1}r_{3}-r_{2}r_{4}) \right\}
\nonumber \\ & & \hspace{0.5in} \mbox{}
+\kappa_{+}^{2} \left\{ \left[ d (m^{2}-a^{2}-e^{2}) - c^{2}
+ a(ab+ce) \right] (r_{1}r_{4}+r_{2}r_{3})
\nonumber \right. \\ & & \hspace{0.5in} \left. \mbox{}
+ i\kappa_{-} (ab+ce-ad) (r_{2}r_{3}-r_{1}r_{4}) \right\}
\nonumber \\ & & \hspace{0.5in} \mbox{}
-4d \left[ b(m^{2}-e^{2}) + c(ae+c) \right] (r_{2}r_{1}+r_{4}r_{3})
\label{du} \; , \\
\D V & = & \kappa_{+}\kappa_{-} \left\{ d [e (m^{2}-a^{2}-e^{2})- 2ac]
(r_{4}+r_{3}-r_{2}-r_{1})
+ de\kappa_{+}\kappa_{-} (r_{2}+r_{1}+r_{4}+r_{3})
\nonumber \right. \\ & & \hspace{0.5in} \mbox{}
+ icd \left[(\kappa_{+}+\kappa_{-})(r_{2}-r_{1})
 + (\kappa_{+}-\kappa_{-})(r_{3}-r_{4}) \right]
\nonumber \\ & & \hspace{0.5in} \left. \mbox{}
+ i [e(ab+ce)+bc] \left[(\kappa_{+}+\kappa_{-})(r_{3}-r_{4})
  + (\kappa_{+}-\kappa_{-})(r_{2}-r_{1}) \right] \right\}
\label{dv} \; , \\
\D W & = & m \kappa_{+}\kappa_{-} \left\{ d \left[ (m^{2}-a^{2}-e^{2})
(r_{4}+r_{3}-r_{2}-r_{1}) + \kappa_{+}\kappa_{-} (r_{2}+r_{1}+r_{4}+r_{3})
\right]
\right. \nonumber \\ & & \hspace{0.5in} \left. \mbox{}
+ i (ab+ce) \left[(\kappa_{+}+\kappa_{-})(r_{3}-r_{4})
	+ (\kappa_{+}-\kappa_{-})(r_{2}-r_{1}) \right] \right\} \; .
\label{dw}
\end{eqnarray}
\end{mathletters}
Of course, there are similar expressions for $(\D U)^{*}:=\D^{*}U^{*}$,
etc.

The real cross sections of the complexified spacetime are easily
identified.  They correspond to real values of $\kappa_{+}^{2}$
and $\kappa_{-}^{2}$.  The solution given in Eqs.\ (\ref{du}),
(\ref{dv}) and (\ref{dw}) is valid not only when $0 > \kappa_{+}^2
> \kappa_{-}^{2}$, but for other values of $\kappa_{+}$ and $\kappa_{-}$
as well.  When both $\kappa_{+}$ and $\kappa_{-}$ are real, the solution
is identical to the five-parameter electrovac solution published recently
by Manko et al.\cite{Manko} in which the parameters $m,a,b,e,c$ were
associated, respectively, with the mass, the rotation, the mass
quadrupole moment, the electric charge and the magnetic dipole moment.

It should be observed that $(\D U)^{*}$, $(\D V)^{*}$ and $(\D W)^{*}$
have the same functional form if $\kappa_{\pm}$ and $r_{a} \;
(a=1,2,3,4)$ are treated as real as they have if $\kappa_{\pm}$ are
treated as imaginary, with $r_{1}^{*}=r_{2}$ and $r_{3}^{*}=r_{4}$.
This means that the expression obtained by Manko et al.\ for the
metric fields $f$, $\gamma$ and $\omega$ in
\begin{equation}
ds^{2} = f^{-1} \left\{ e^{2\gamma} (dz^{2}+d\rho^{2})
+ \rho^{2} d\varphi^{2} \right\} - f (dt-\omega d\varphi)^{2} \; .
\end{equation}
will hold for the other cases as well.  In a later paper concerned with
the complexified Cosgrove transformation, we shall develop a completely
general formula for the field $\omega$.  At this time, we merely remark
that the field $\gamma$ is given by\cite{referee}
\begin{equation}
e^{2\gamma} = \frac{|U|^{2}+|V|^{2}-|W|^{2}}
{|U_{0}|^{2} r_{1}r_{2}r_{3}r_{4}} \; ,
\end{equation}
and the field $f$ is given by
\begin{equation}
f = \Re{\E} + |\Phi|^{2} = \frac{|U|^{2}+|V|^{2}-|W|^{2}}{|U+W|^{2}}
\; .
\end{equation}
Thus, the infinite red shift surface corresponds to
\begin{equation}
|U|^{2}+|V|^{2}-|W|^{2} = 0 \; ,
\end{equation}
and the curvature singularities occur at $U+W=0$.

\section{Toward a purely algebraic derivation}

It was Kinnersley\cite{Kin} who first pointed out that $U$, $V$ and $W$
could always be selected so that the field equations
\begin{equation}
(|U|^{2}+|V|^{2}-|W|^{2}) \; \nabla^{2} \left( \begin{array}{c}
U \\ V \\ W
\end{array} \right) = 2 (U^{*} \nabla U + V^{*} \nabla V
- W^{*} \nabla W) \cdot \nabla \left( \begin{array}{c}
U \\ V \\ W
\end{array} \right) \label{kin}
\end{equation}
are satisfied.  The reader will find it instructive to work out the
$n=1$ solution of these equations, where
\begin{mathletters}
\begin{eqnarray}
U & = & \sum_{i} u_{i} r_{i} \; , \\
V & = & v \; , \\
W & = & w \; ,
\end{eqnarray}
\end{mathletters}
and $u_{1}$, $u_{2}$, $v$ and $w$ are complex constants.  This is
not difficult to do, if one observes that
\begin{equation}
\nabla^{2} r_{i} = \frac{2}{r_{i}} \; , \label{D2}
\end{equation}
and
\begin{equation}
\nabla r_{i} \cdot \nabla r_{j} =
\frac{r_{i}^{2}+r_{j}^{2}-(K_{i}-K_{j})^{2}}{2r_{i}r_{j}} \; ,
\label{D1D1}
\end{equation}
and one uses the relation\cite{MS}
\begin{equation}
|U(z,0)|^{2}+|V(z,0)|^{2}-|W(z,0)|^{2} = |U_{0}|^{2}
\Pi_{a=1}^{2n} (z-K_{a}) \; .
\label{quad}
\end{equation}
In the present paper we have been interested in $n=2$ solutions, in which
\begin{mathletters}
\begin{eqnarray}
U & = & \sum_{i<j} u_{ij} r_{i} r_{j} \; , \\
V & = & \sum_{i} v_{i} r_{i} \; , \\
W & = & \sum_{i} w_{i} r_{i} \; ,
\end{eqnarray}
\end{mathletters}
where the $u_{ij}$, the $v_{i}$ and the $w_{i}$ are complex constants.
For all values of $n$ the mechanism of solution remains the same as that
illustrated by the $n=1$ case, but the algebra becomes increasingly more
difficult as $n$ increases.

It would be nice if one could formulate a simple strictly algebraic
derivation of the general solution of Eqs.\ (\ref{kin}) corresponding
to rational axis data by using Eqs.\ (\ref{D2}), (\ref{D1D1}) and
(\ref{quad}).  We shall postpone further consideration of this approach
until a later paper, where we shall be concerned primarily with $n>2$.

\section*{Acknowledgement}
This work was supported in part by grant PHY-93-07762 from the
National Science Foundation.

\end{document}